\documentclass{article}

\usepackage{times}
\usepackage{graphicx} 
\usepackage{subfigure}

\usepackage{natbib}

\usepackage{algorithm}
\usepackage{algorithmic}
\usepackage[algo2e,linesnumbered,lined,boxed,vlined]{algorithm2e}

\usepackage{amsfonts}
\usepackage{amssymb}

\usepackage{hyperref}

\usepackage[accepted]{icml2016}

\usepackage{comment}
\usepackage{amsthm}
\usepackage{mathtools}
\usepackage{tabularx}
\usepackage{multirow}

\def\b{\ensuremath\boldsymbol}

\usepackage{lastpage}
\usepackage{fancyhdr}
\icmltitlerunning{Fitting A Mixture Distribution to Data: Tutorial}

\begin{document}

\twocolumn[
\icmltitle{Fitting A Mixture Distribution to Data: Tutorial}

\icmlauthor{Benyamin Ghojogh}{bghojogh@uwaterloo.ca}
\icmladdress{Department of Electrical and Computer Engineering, 
\\Machine Learning Laboratory, University of Waterloo, Waterloo, ON, Canada}
\icmlauthor{Aydin Ghojogh}{aydin.ghojogh@gmail.com}
\icmladdress{}
\icmlauthor{Mark Crowley}{mcrowley@uwaterloo.ca}
\icmladdress{Department of Electrical and Computer Engineering, 
\\Machine Learning Laboratory, University of Waterloo, Waterloo, ON, Canada}
\icmlauthor{Fakhri Karray}{karray@uwaterloo.ca}
\icmladdress{Department of Electrical and Computer Engineering, 
\\Centre for Pattern Analysis and Machine Intelligence, University of Waterloo, Waterloo, ON, Canada}

\icmlkeywords{boring formatting information, machine learning, ICML}

\vskip 0.3in
]

\begin{abstract}
This paper is a step-by-step tutorial for fitting a mixture distribution to data. It merely assumes the reader has the background of calculus and linear algebra. Other required background is briefly reviewed before explaining the main algorithm. In explaining the main algorithm, first, fitting a mixture of two distributions is detailed and examples of fitting two Gaussians and Poissons, respectively for continuous and discrete cases, are introduced. Thereafter, fitting several distributions in general case is explained and examples of several Gaussians (Gaussian Mixture Model) and Poissons are again provided. Model-based clustering, as one of the applications of mixture distributions, is also introduced. Numerical simulations are also provided for both Gaussian and Poisson examples for the sake of better clarification. 
\end{abstract}

\section{Introduction}

Every random variable can be considered as a sample from a distribution, whether a well-known distribution or a not very well-known (or ``ugly'') distribution. Some random variables are drawn from one single distribution, such as a normal distribution. But life is not always so easy! Most of real-life random variables might have been generated from a mixture of several distributions and not a single distribution. The mixture distribution is a weighted summation of $K$ distributions $\{g_1(x; \Theta_1), \dots, g_K(x; \Theta_K)\}$ where the weights $\{w_1, \dots, w_K\}$ sum to one. As is obvious, every distribution in the mixture has its own parameter $\Theta_k$. The mixture distribution is formulated as:
\begin{equation}\label{equation_mixture}
\begin{aligned}
& f(x; \Theta_1, \dots, \Theta_K) = \sum_{k=1}^K w_k\, g_k(x; \Theta_k), \\
& \text{subject to} ~~~~ \sum_{k=1}^K w_k = 1.
\end{aligned}
\end{equation}
The distributions can be from different families, for example from beta and normal distributions. However, this makes the problem very complex and sometimes useless; therefore, mostly the distributions in a mixture are from one family (e.g., all normal distributions) but with different parameters. 
This paper aims to find the parameters of the distributions in the mixture distribution $f(x;\Theta)$ as well as the weights (also called ``mixing probabilities'') $w_k$.

The remainder of paper is organized as follows. Section \ref{section_background} reviews some technical background required for explaining the main algorithm. Afterwards, the methodology of fitting a mixture distribution to data is explained in Section \ref{section_mixture_distibution}. In that section, first the mixture of two distributions, as a special case of mixture distributions, is introduced and analyzed. Then, the general mixture distribution is discussed. Meanwhile, examples of mixtures of Gaussians (example for continuous cases) and Poissons (example for discrete cases) are mentioned for better clarification. 
Section \ref{section_clustering} briefly introduces clustering as one of the applications of mixture distributions. 
In Section \ref{section_simulations}, the discussed methods are then implemented through some simulations in order to have better sense of how these algorithms work. Finally, Section \ref{section_conclusions} concludes the paper.

\section{Background}\label{section_background}

This section reviews some technical background required for explaining the main algorithm. This review includes probability and Bayes rule, probability mass/density function, expectation, maximum likelihood estimation, expectation maximization, and Lagrange multiplier.

\subsection{Probability and Bayes Rule}

If $S$ denotes the total sample space and $A$ denotes an event in this sample space, the probability of event $A$ is:
\begin{align}
\mathbb{P}(A) = \frac{|A|}{|S|}.
\end{align}
The conditional probability, i.e., probability of occurance of event $A$ given that event $B$ happens, is:
\begin{align}
\mathbb{P}(A|B) & = \frac{\mathbb{P}(A,B)}{\mathbb{P}(B)} \label{equation_Bayes_1}\\
& = \frac{\mathbb{P}(B|A)\,\mathbb{P}(A)}{\mathbb{P}(B)}, \label{equation_Bayes_2}
\end{align}
where $\mathbb{P}(A|B)$, $\mathbb{P}(B|A)$, $\mathbb{P}(A)$, and $\mathbb{P}(B)$ are called likelihood, posterior, prior, and marginal probabilities, respectively. 
If we assume that the event $A$ consists of some cases $A=\{A_1, \dots, A_n\}$, we can write:
\begin{align} \label{equation_Bayes_3}
\mathbb{P}(A_i|B) = \frac{\mathbb{P}(B|A_i)\,\mathbb{P}(A_i)}{\sum_{j=1}^n \mathbb{P}(B|A_j)\,\mathbb{P}(A_j)}.
\end{align}
the equations (\ref{equation_Bayes_2}) and (\ref{equation_Bayes_3}) are two versions of Bayes rule.

\subsection{Probability Mass/Density Function}

In discrete cases, the probability mass function is defined as:
\begin{align}
f(x) = \mathbb{P}(X=x), 
\end{align}
where $X$ and $x$ are a random variable and a number, respectively. 

In continuous cases, the probability density function is:
\begin{align}
f(x) = \lim_{\Delta x \rightarrow 0} \frac{\mathbb{P}(x \leq X \leq x + \Delta x)}{\Delta x} = \frac{\partial \mathbb{P}(X \leq x)}{\partial x}.
\end{align}
In this work, by mixture of distributions, we imply mixture of mass/density functions.

\subsection{Expectation}

Expectation means the value of a random variable $X$ on average. Therefore, expectation is a weighted average where the weights are probabilities of the random variable $X$ to get different values. In discrete and continuous cases, the expectation is:
\begin{align}
& \mathbb{E}(X) = \sum_{\textbf{dom } x} x f(x), \\
& \mathbb{E}(X) = \int\displaylimits_{\textbf{dom } x} x f(x) \,dx,
\end{align}
respectively, where $\textbf{dom } x$ is the domain of $X$.
The conditional expectation is defined as:
\begin{align}
& \mathbb{E}_{X|Y}(X|Y) = \sum_{\textbf{dom } x} x f(x|y), \\
& \mathbb{E}_{X|Y}(X|Y) = \int\displaylimits_{\textbf{dom } x} x f(x|y) \,dx,
\end{align}
for discrete and continuous cases, respectively. 

\subsection{Maximum Likelihood Estimation}

Assume we have a sample with size $n$, i.e., $\{x_1, \dots, x_n\}$. Also assume that we know the distribution from which this sample has been randomly drawn but we do not know the parameters of that distribution. For example, we know it is drawn from a normal distribution but the mean and variance of this distribution are unknown. The goal is to estimate the parameters of the distribution using the sample $\{x_1, \dots, x_n\}$ available from it. This estimation of parameters from the available sample is called ``point estimation''. One of the approaches for point estimation is Maximum Likelihood Estimation (MLE). As it is obvious from its name, MLE deals with the likelihood of data.

We postulate that the values of sample, i.e., $x_1, \dots, x_n$, are independent random variates of data having the sample distribution. In other words, the data has a joint distribution $f_X(x_1, \dots, x_n | \Theta)$ with parameter $\Theta$ and we assume the variates are independent and identically distributed ($iid$) variates, i.e., $x_i \overset{iid}{\sim} f_X(x_i; \Theta)$ with the same parameter $\Theta$.
Considering the Bayes rule, equation (\ref{equation_Bayes_2}), we have:
\begin{align}
L(\Theta| x_1, \dots, x_n) = \frac{f_X(x_1, \dots, x_n | \Theta) \pi(\Theta)}{f_X(x_1, \dots, x_n)}.
\end{align}
The MLE aims to find parameter $\Theta$ which maximizes the likelihood:
\begin{align}
\widehat{\Theta} = \arg \max_\Theta L(\Theta).
\end{align}
According to the definition, the likelihood can be written as:
\begin{align}
L(\Theta | x_1, \dots, x_n) & := f(x_1, \dots, x_n;\Theta) \nonumber \\
& \overset{(a)}{=} \prod_{i=1}^n f(x_i, \Theta),
\end{align}
where $(a)$ is because the $x_1, \dots, x_n$ are $iid$.
Note that in literature, the $L(\Theta | x_1, \dots, x_n)$ is also denoted by $L(\Theta)$ for simplicity.

Usually, for more convenience, we use log-likelihood rather than likelihood:
\begin{align}
\ell(\Theta) & := \log L(\Theta) \label{equation_log_likelihood} \\
& = \log \prod_{i=1}^n f(x_i, \Theta) = \sum_{i=1}^n \log f(x_i, \Theta).
\end{align}
Often, the logarithm is a natural logarithm for the sake of compatibility with the exponential in the well-known normal density function. 
Notice that as logarithm function is monotonic, it does not change the location of maximization of the likelihood. 

\subsection{Expectation Maximization}

Sometimes, the data are not fully observable. For example, the data are known to be whether zero or greater than zero. As an illustration, assume the data are collected for a particular disease but for convenience of the patients participated in the survey, the severity of the disease is not recorded but only the existence or non-existence of the disease is reported. So, the data are not giving us complete information as $X_i>0$ is not obvious whether is $X_i=2$ or $X_i=1000$.

In this case, MLE cannot be directly applied as we do not have access to complete information and some data are missing.
In this case, Expectation Maximization (EM) is useful. 
The main idea of EM can be summarized in this short friendly conversation:

\begin{itshape}
-- What shall we do? The data is missing! The log-likelihood is not known completely so MLE cannot be used.
\newline
-- Mmm, probably we can replace the missing data with something...
\newline
-- Aha! Let us replace it with its mean. 
\newline
-- You are right! We can take the mean of log-likelihood over the possible values of the missing data. Then everything in the log-likelihood will be known, and then...
\newline
-- And then we can do MLE!
\end{itshape}

Assume $D^{(obs)}$ and $D^{(miss)}$ denote the observed data ($X_i$'s $=0$ in the above example) and the missing data ($X_i$'s $>0$ in the above example).
The EM algorithm includes two main steps, i.e., E-step and M-step.

In the E-step, the log-likelihood (equation (\ref{equation_log_likelihood})), is taken expectation with respect to the missing data $D^{(miss)}$ in order to have a mean estimation of it. Let $Q(\Theta)$ denote the expectation of the likelihood with respect to $D^{(miss)}$:
\begin{align}
Q(\Theta) := \mathbb{E}_{D^{(miss)}|D^{(obs)}, \Theta} [\ell(\Theta)].
\end{align}
Note that in the above expectation, the $D^{(obs)}$ and $\Theta$ are conditioned on, so they are treated as constants and not random variables.

In the M-step, the MLE approach is used where the log-likelihood is replaced with its expectation, i.e., $Q(\Theta)$; therefore:
\begin{align}
\widehat{\Theta} = \arg \max_\Theta Q(\Theta).
\end{align}
These two steps are iteratively repeated until convergence of the estimated parameters $\widehat{\Theta}$.

\subsection{Lagrange Multiplier}\label{section_lagrangeMultiplier}

Suppose we have a multi-variate function $Q(\Theta_1, \dots, \Theta_K)$ (called ``objective function'') and we want to maximize (or minimize) it. However, this optimization is constrained and its constraint is equality $P(\Theta_1, \dots, \Theta_K) = c$ where $c$ is a constant. So, the constrained optimization problem is:
\begin{equation}
\begin{aligned}
& \underset{\Theta_1, \dots, \Theta_K}{\text{maximize}}
& & Q(\Theta_1, \dots, \Theta_K), \\
& \text{subject to}
& & P(\Theta_1, \dots, \Theta_K) = c.
\end{aligned}
\end{equation}
For solving this problem, we can introduce a new variable $\alpha$ which is called ``Lagrange multiplier''. Also, a new function $\mathcal{L}(\Theta_1, \dots, \Theta_K, \alpha)$, called ``Lagrangian'' is introduced:
\begin{equation}
\begin{aligned}
\mathcal{L}(\Theta_1, \dots, \Theta_K, \alpha) = &\, Q(\Theta_1, \dots, \Theta_K) \\
&- \alpha \big(P(\Theta_1, \dots, \Theta_K) - c\big).
\end{aligned}
\end{equation}
Maximizing (or minimizing) this Lagrangian function gives us the solution to the optimization problem \cite{boyd2004convex}:
\begin{align}
\nabla_{\Theta_1, \dots, \Theta_K, \alpha} \mathcal{L} \overset{\text{set}}{=} 0,
\end{align}
which gives us:
\begin{align*}
\nabla_{\Theta_1, \dots, \Theta_K} \mathcal{L} \overset{\text{set}}{=} 0 &\implies \nabla_{\Theta_1, \dots, \Theta_K} Q = \alpha \nabla_{\Theta_1, \dots, \Theta_K} P, \\
\nabla_{\alpha} \mathcal{L} \overset{\text{set}}{=} 0 & \implies P(\Theta_1, \dots, \Theta_K) = c.
\end{align*}

\section{Fitting A Mixture Distribution}\label{section_mixture_distibution}

As was mentioned in the introduction, the goal of fitting a mixture distribution is to find the parameters and weights of a weighted summation of distributions (see equation (\ref{equation_mixture})).
First, as a spacial case of mixture distributions, we work on mixture of two distributions and then we discuss the general mixture of distributions. 

\subsection{Mixture of Two Distributions}

Assume that we want to fit a mixture of two distributions $g_1(x; \Theta_1)$ and $g_2(x; \Theta_2)$ to the data. Note that, in theory, these two distributions are not necessarily from the same distribution family. As we have only two distributions in the mixture, equation (\ref{equation_mixture}) is simplified to:
\begin{equation}\label{equation_twoMixture_general}
\begin{aligned}
f(x; \Theta_1, \Theta_2) = w\, g_1(x; \Theta_1) + (1-w)\, g_2(x; \Theta_2).
\end{aligned}
\end{equation}
Note that the parameter $w$ (or $w_k$ in general) is called ``mixing probability'' \cite{friedman2001elements} and is sometimes denoted by $\pi$ (or $\pi_k$ in general) in literature. 

The likelihood and log-likelihood for this mixture is:
\begin{alignat*}{2}
& L(\Theta_1, \Theta_2) && = f(x_1, \dots, x_n; \Theta_1, \Theta_2) \overset{(a)}{=} \prod_{i=1}^n f(x_i; \Theta_1, \Theta_2) \nonumber \\
& && = \prod_{i=1}^n \Big[w\, g_1(x_i; \Theta_1) + (1-w)\, g_2(x_i; \Theta_2)\Big], 
\end{alignat*}
\begin{align*}
\ell(\Theta_1, \Theta_2) = \sum_{i=1}^n \log \Big[&w\, g_1(x_i; \Theta_1) \\
&+ (1-w)\, g_2(x_i; \Theta_2)\Big],
\end{align*}
where $(a)$ is because of the assumption that $x_1, \dots, x_n$ are $iid$.
Optimizing this log-likelihood is difficult because of the summation within the logarithm. However, we can use a nice trick here \cite{friedman2001elements}: Let $\Delta_i$ be defined as:
\begin{align*}
\Delta_i := 
\left\{
    \begin{array}{ll}
        1 ~~~ & \text{if } x_i \text{ belongs to } g_1(x; \Theta_1), \\
        0 ~~~ & \text{if } x_i \text{ belongs to } g_2(x; \Theta_2),
     \end{array}
\right.
\end{align*}
and its probability be:
\begin{align*}
\left\{
    \begin{array}{ll}
        \mathbb{P}(\Delta_i = 1) = w, \\
        \mathbb{P}(\Delta_i = 0) = 1-w.
     \end{array}
\right.
\end{align*}
Therefore, the log-likelihood can be written as:
\begin{align*}
& \ell(\Theta_1, \Theta_2) = \\
& \left\{
    \begin{array}{ll}
        \sum_{i=1}^n \log \big[ w\, g_1(x_i; \Theta_1) \big]  & \text{if } \Delta_i = 1\\\\
        \sum_{i=1}^n \log \big[ (1-w)\, g_2(x_i; \Theta_2) \big] & \text{if } \Delta_i = 0
     \end{array}
\right.
\end{align*}
The above expression can be restated as:
\begin{align*}
\ell(\Theta_1, \Theta_2) = \sum_{i=1}^n \Big[ & \Delta_i \log \big[ w\, g_1(x_i; \Theta_1) \big] + \\
& (1-\Delta_i) \log \big[ (1-w)\, g_2(x_i; \Theta_2) \big] \Big].
\end{align*}
The $\Delta_i$ here is the incomplete (missing) datum because we do not know whether it is $\Delta_i=0$ or $\Delta_i=1$ for $x_i$. Hence, using the EM algorithm, we try to estimate it by its expectation. 

The E-step in EM:
\begin{align*}
Q(\Theta_1,& \Theta_2) = \sum_{i=1}^n \Big[  \mathbb{E}[\Delta_i|X,\Theta_1, \Theta_2] \log \big[ w\, g_1(x_i; \Theta_1) \big] + \\
& \mathbb{E}[(1-\Delta_i)|X,\Theta_1, \Theta_2] \log \big[ (1-w)\, g_2(x_i; \Theta_2) \big] \Big].
\end{align*}
Notice that the above expressions are linear with respect to $\Delta_i$ and that is why the two logarithms were factored out. Assume $\widehat{\gamma}_i := \mathbb{E}[\Delta_i|X, \Theta_1, \Theta_2]$ which is called ``responsibility'' of $x_i$ \cite{friedman2001elements}.

The $\Delta_i$ is either $0$ or $1$; therefore:
\begin{alignat*}{2}
\mathbb{E}[\Delta_i|X, \Theta_1, \Theta_2] &= \,&& 0 \times \mathbb{P}(\Delta_i=0|X, \Theta_1, \Theta_2) + \\
& && 1 \times \mathbb{P}(\Delta_i=1|X, \Theta_1, \Theta_2) \\
& = &&\mathbb{P}(\Delta_i=1|X, \Theta_1, \Theta_2).
\end{alignat*}
According to Bayes rule (equation (\ref{equation_Bayes_3})), we have:
\begin{align*}
\mathbb{P}(\Delta_i&=1 |X, \Theta_1, \Theta_2) \\
& = \frac{\mathbb{P}(X, \Theta_1, \Theta_2, \Delta_i=1)}{\mathbb{P}(X; \Theta_1, \Theta_2)} \\
& = \frac{\mathbb{P}(X, \Theta_1, \Theta_2| \Delta_i=1)\,\mathbb{P}(\Delta_i=1)}{\sum_{j=0}^1 \mathbb{P}(X, \Theta_1, \Theta_2| \Delta_i=j)\,\mathbb{P}(\Delta_i=j)}.
\end{align*}
The marginal probability in the denominator is:
\begin{align*}
\mathbb{P}(X; \Theta_1, \Theta_2) & = (1-w)\, g_2(x_i;\Theta_2) + w\, g_1(x_i;\Theta_1).
\end{align*}
Thus:
\begin{align}\label{equation_gamma_twoMixture}
\widehat{\gamma}_i = \frac{\widehat{w}\, g_1(x_i;\Theta_1)}{\widehat{w}\, g_1(x_i;\Theta_1) + (1-\widehat{w})\, g_2(x_i;\Theta_2)},
\end{align}
and 
\begin{equation}
\begin{aligned}
Q(\Theta_1, \Theta_2) = & \sum_{i=1}^n \Big[   \widehat{\gamma}_i \log \big[ w\, g_1(x_i; \Theta_1) \big] + \\
& (1-\widehat{\gamma}_i) \log \big[ (1-w)\, g_2(x_i; \Theta_2) \big] \Big].
\end{aligned}
\end{equation}
Some simplification of $Q(\Theta_1, \Theta_2)$ will help in next step:
\begin{equation*}
\begin{aligned}
Q(\Theta_1, &\Theta_2) = \sum_{i=1}^n \Big[   \widehat{\gamma}_i \log w + \widehat{\gamma}_i \log g_1(x_i; \Theta_1) + \\
& (1-\widehat{\gamma}_i) \log (1-w) + (1-\widehat{\gamma}_i) \log g_2(x_i; \Theta_2) \Big].
\end{aligned}
\end{equation*}

The M-step in EM:
\begin{align*}
\widehat{\Theta}_1, \widehat{\Theta}_2, \widehat{w} = \arg \max_{\Theta_1, \Theta_2, w} Q(\Theta_1, \Theta_2, w).
\end{align*}
Note that the function $Q(\Theta_1, \Theta_2)$ is also a function of $w$ and that is why we wrote it as $Q(\Theta_1, \Theta_2, w)$.
\begin{align}
& \frac{\partial Q}{\partial \Theta_1} = \sum_{i=1}^n \Big[\frac{\widehat{\gamma}_i}{g_1(x_i;\Theta_1)} \frac{\partial g_1(x_i;\Theta_1)}{\partial \Theta_1} \Big] \overset{\text{set}}{=} 0, \label{equation_derivative_theta1_twoMixture} \\
& \frac{\partial Q}{\partial \Theta_2} = \sum_{i=1}^n \Big[\frac{1-\widehat{\gamma}_i}{g_2(x_i;\Theta_1)} \frac{\partial g_2(x_i;\Theta_2)}{\partial \Theta_2} \Big] \overset{\text{set}}{=} 0, \label{equation_derivative_theta2_twoMixture} \\
& \frac{\partial Q}{\partial w} = \sum_{i=1}^n \Big[\widehat{\gamma}_i (\frac{1}{w}) + (1-\widehat{\gamma}_i)(\frac{-1}{1-w})\Big] \overset{\text{set}}{=} 0, \nonumber \\
& \implies \widehat{w} = \frac{1}{n} \sum_{i=1}^n \widehat{\gamma}_i \label{equation_w_twoMixture}
\end{align}
So, the mixing probability is the average of the responsibilities which makes sense. Solving equations (\ref{equation_derivative_theta1_twoMixture}), (\ref{equation_derivative_theta2_twoMixture}), and (\ref{equation_w_twoMixture}) gives us the estimations $\widehat{\Theta}_1$, $\widehat{\Theta}_2$, and $\widehat{w}$ in every iteration.

The iterative algorithm for finding the parameters of the mixture of two distributions is shown in Algorithm \ref{algorithm_twoMixture}.

\SetAlCapSkip{0.5em}
\IncMargin{0.8em}
\begin{algorithm2e}[!t]
\DontPrintSemicolon
    \textbf{START:}
	\textbf{Initialize} $\widehat{\Theta}_1$, $\widehat{\Theta}_2$, $\widehat{w}$\;
	\While{not converged}{
	    \textit{// E-step in EM:}\;
	    \For{$i$ from $1$ to $n$}{
            $\widehat{\gamma}_i  \gets $ equation (\ref{equation_gamma_twoMixture})\;
        }
        \textit{// M-step in EM:}\;
	    $\widehat{\Theta}_1  \gets $ equation (\ref{equation_derivative_theta1_twoMixture})\;
	    $\widehat{\Theta}_2  \gets $ equation (\ref{equation_derivative_theta2_twoMixture})\;
	    $\widehat{w} \gets $ equation (\ref{equation_w_twoMixture})\;
	    \textit{// Check convergence:}\;
	    Compare $\widehat{\Theta}_1$, $\widehat{\Theta}_2$, and $\widehat{w}$ with their values in previous iteration
	}
\caption{Fitting A Mixture of Two Distributions}\label{algorithm_twoMixture}
\end{algorithm2e}
\DecMargin{0.8em}

\subsubsection{Mixture of Two Gaussians}

Here, we consider a mixture of two one-dimensional Gaussian distributions as an example for mixture of two continuous distributions. In this case, we have:
\begin{alignat*}{2}
& g_1(x; \mu_1, \sigma_1^2) &&= \frac{1}{\sqrt{2\pi\sigma_1^2}} \exp (-\frac{(x-\mu_1)^2}{2\sigma_1^2}) \\
& &&= \phi(\frac{x-\mu_1}{\sigma_1}),\\
& g_2(x; \mu_2, \sigma_2^2) &&= \frac{1}{\sqrt{2\pi\sigma_2^2}} \exp (-\frac{(x-\mu_2)^2}{2\sigma_2^2}) \\
& &&= \phi(\frac{x-\mu_2}{\sigma_2}),
\end{alignat*}
where $\phi(x)$ is the probability density function of normal distribution.
Therefore, equation (\ref{equation_twoMixture_general}) becomes:
\begin{equation}\label{equation_twoGaussians_general}
\begin{aligned}
f(x; \mu_1, \mu_2,&\, \sigma_1^2, \sigma_2^2) = \\ 
& w\, \phi(\frac{x-\mu_1}{\sigma_1}) + (1-w)\, \phi(\frac{x-\mu_2}{\sigma_2}).
\end{aligned}
\end{equation}
The equation (\ref{equation_gamma_twoMixture}) becomes: 
\begin{align}\label{equation_gamma_twoGaussians}
\widehat{\gamma}_i = \frac{\widehat{w}\, \phi(\frac{x_i-\mu_1}{\sigma_1})}{\widehat{w}\, \phi(\frac{x_i-\mu_1}{\sigma_1}) + (1-\widehat{w})\,\phi(\frac{x_i-\mu_2}{\sigma_2})}.
\end{align}
The $Q(\mu_1, \mu_2, \sigma_1^2, \sigma_2^2)$ is:
\begin{alignat*}{2}
Q(\mu_1,&\, \mu_2, \sigma_1^2, \sigma_2^2) = \sum_{i=1}^n \Big[   \widehat{\gamma}_i \log w 
\\& + \widehat{\gamma}_i\, (-\frac{1}{2} \log (2\pi) - \log \sigma_1 -\frac{(x_i-\mu_1)^2}{2\sigma_1^2})\\
& +(1-\widehat{\gamma}_i) \log (1-w) \\
& + (1-\widehat{\gamma}_i) (-\frac{1}{2} \log (2\pi) - \log \sigma_2 -\frac{(x_i-\mu_2)^2}{2\sigma_2^2}) \Big].
\end{alignat*}
Therefore:
\begin{align}
& \frac{\partial Q}{\partial \mu_1} = \sum_{i=1}^n \Big[\widehat{\gamma}_i\, (\frac{x_i-\mu_1}{\sigma_1^2}) \Big] \overset{\text{set}}{=} 0, \nonumber \\
& \implies \widehat{\mu}_1 = \frac{\sum_{i=1}^n \widehat{\gamma}_i\, x_i}{\sum_{i=1}^n \widehat{\gamma}_i}, \label{equation_mu1_twoGaussians}\\
& \frac{\partial Q}{\partial \mu_2} = \sum_{i=1}^n \Big[(1-\widehat{\gamma}_i) (\frac{x_i-\mu_2}{\sigma_2^2}) \Big] \overset{\text{set}}{=} 0, \nonumber \\
& \implies \widehat{\mu}_2 = \frac{\sum_{i=1}^n (1-\widehat{\gamma}_i)\, x_i}{\sum_{i=1}^n (1-\widehat{\gamma}_i)}, \label{equation_mu2_twoGaussians} \\
& \frac{\partial Q}{\partial \sigma_1} = \sum_{i=1}^n \Big[\widehat{\gamma}_i\, (\frac{-1}{\sigma_1} + \frac{(x_i - \mu_1)^2}{\sigma_1^3}) \Big] \overset{\text{set}}{=} 0, \nonumber \\
& \implies \widehat{\sigma}_1^2 = \frac{\sum_{i=1}^n \widehat{\gamma}_i\, (x_i-\widehat{\mu}_1)^2}{\sum_{i=1}^n \widehat{\gamma}_i}, \label{equation_sigma1_twoGaussians}\\
& \frac{\partial Q}{\partial \sigma_2} = \sum_{i=1}^n \Big[(1-\widehat{\gamma}_i) (\frac{-1}{\sigma_2} + \frac{(x_i - \mu_2)^2}{\sigma_2^3}) \Big] \overset{\text{set}}{=} 0, \nonumber \\
& \implies \widehat{\sigma}_2^2 = \frac{\sum_{i=1}^n (1-\widehat{\gamma}_i) (x_i-\widehat{\mu}_2)^2}{\sum_{i=1}^n (1-\widehat{\gamma}_i)}, \label{equation_sigma2_twoGaussians}
\end{align}
and $\widehat{w}$ is the same as equation (\ref{equation_w_twoMixture}).

Iteratively solving equations (\ref{equation_gamma_twoGaussians}), (\ref{equation_mu1_twoGaussians}), (\ref{equation_mu2_twoGaussians}), (\ref{equation_sigma1_twoGaussians}), (\ref{equation_sigma2_twoGaussians}), and (\ref{equation_w_twoMixture}) using Algorithm (\ref{algorithm_twoMixture}) gives us the estimations for $\widehat{\mu}_1$, $\widehat{\mu}_2$, $\widehat{\sigma}_1$, $\widehat{\sigma}_2$, and $\widehat{w}$ in equation (\ref{equation_twoGaussians_general}).

\subsubsection{Mixture of Two Poissons}

Here, we consider a mixture of two Poisson distributions as an example for mixture of two discrete distributions. In this case, we have:
\begin{align*}
& g_1(x; \lambda_1) = \frac{e^{-\lambda_1} \lambda_1^{x}}{x!}, \\
& g_2(x; \lambda_2) = \frac{e^{-\lambda_2} \lambda_2^{x}}{x!},
\end{align*}
therefore, equation (\ref{equation_twoMixture_general}) becomes:
\begin{equation}\label{equation_twoPoissons_general}
\begin{aligned}
f(x; \lambda_1, \lambda_2) = w\, \frac{e^{-\lambda_1} \lambda_1^{x}}{x!} + (1-w)\, \frac{e^{-\lambda_2} \lambda_2^{x}}{x!}.
\end{aligned}
\end{equation}
The equation (\ref{equation_gamma_twoMixture}) becomes: 
\begin{align}\label{equation_gamma_twoPoissons}
\widehat{\gamma}_i = \frac{\widehat{w}\, (\frac{e^{-\widehat{\lambda}_1} \widehat{\lambda}_1^{x_i}}{x_i!})}{\widehat{w}\, (\frac{e^{-\widehat{\lambda}_1} \widehat{\lambda}_1^{x_i}}{x_i!}) + (1-\widehat{w})\,(\frac{e^{-\widehat{\lambda}_2} \widehat{\lambda}_2^{x_i}}{x_i!})}.
\end{align}
The $Q(\lambda_1, \lambda_2)$ is:
\begin{alignat*}{2}
Q(\lambda_1, \lambda_2) = &\sum_{i=1}^n \Big[   \widehat{\gamma}_i \log w 
\\& + \widehat{\gamma}_i (-\lambda_1 + x_i \log \lambda_1 - \log x_i!) \\
& +(1-\widehat{\gamma}_i) \log (1-w) \\
& + (1-\widehat{\gamma}_i) (-\lambda_2 + x_i \log \lambda_2 - \log x_i!) \Big].
\end{alignat*}
Therefore:
\begin{align}
& \frac{\partial Q}{\partial \lambda_1} = \sum_{i=1}^n \Big[\widehat{\gamma}_i (-1+\frac{x_i}{\lambda_1}) \Big] \overset{\text{set}}{=} 0, \nonumber \\
& \implies \widehat{\lambda}_1 = \frac{\sum_{i=1}^n \widehat{\gamma}_i\, x_i}{\sum_{i=1}^n \widehat{\gamma}_i}, \label{equation_lambda1_twoPoissons} \\
& \frac{\partial Q}{\partial \lambda_2} = \sum_{i=1}^n \Big[(1-\widehat{\gamma}_i) (-1+\frac{x_i}{\lambda_2}) \Big] \overset{\text{set}}{=} 0, \nonumber \\
& \implies \widehat{\lambda}_2 = \frac{\sum_{i=1}^n (1-\widehat{\gamma}_i)\, x_i}{\sum_{i=1}^n (1-\widehat{\gamma}_i)}, \label{equation_lambda2_twoPoissons}
\end{align}
and $\widehat{w}$ is the same as equation (\ref{equation_w_twoMixture}).

Iteratively solving equations (\ref{equation_gamma_twoPoissons}), (\ref{equation_lambda1_twoPoissons}), (\ref{equation_lambda2_twoPoissons}), and (\ref{equation_w_twoMixture}) using Algorithm (\ref{algorithm_twoMixture}) gives us the estimations for $\widehat{\lambda}_1$, $\widehat{\lambda}_2$, and $\widehat{w}$ in equation (\ref{equation_twoPoissons_general}).

\subsection{Mixture of Several Distributions}

Now, assume a more general case where we want to fit a mixture of $K$ distributions $g_1(x; \Theta_1), \dots, g_K(x; \Theta_K)$ to the data. Again, in theory, these $K$ distributions are not necessarily from the same distribution family. For more convenience of reader, equation (\ref{equation_mixture}) is repeated here:
\begin{equation*}
\begin{aligned}
& f(x; \Theta_1, \dots, \Theta_K) = \sum_{k=1}^K w_k\, g_k(x; \Theta_k), \\
& \text{subject to} ~~~~ \sum_{k=1}^K w_k = 1.
\end{aligned}
\end{equation*}
The likelihood and log-likelihood for this mixture is:
\begin{alignat*}{2}
& L(\Theta_1, \dots, \Theta_K) && = f(x_1, \dots, x_n; \Theta_1, \dots, \Theta_K) \\
& &&\overset{(a)}{=} \prod_{i=1}^n f(x_i; \Theta_1, \dots, \Theta_K) \nonumber \\
& && = \prod_{i=1}^n \sum_{k=1}^K w_k g_k(x_i; \Theta_k)  
\end{alignat*}
\begin{align*}
\ell(\Theta_1, \dots, \Theta_K) = \sum_{i=1}^n \log \Big[\sum_{k=1}^K w_k g_k(x_i; \Theta_k)\Big],
\end{align*}
where $(a)$ is because of assumption that $x_1, \dots, x_n$ are $iid$.
Optimizing this log-likelihood is difficult because of the summation within the logarithm. We use the same trick as the trick mentioned for mixture of two distributions:
\begin{align*}
\Delta_{i,k} := 
\left\{
    \begin{array}{ll}
        1 ~~~ & \text{if } x_i \text{ belongs to } g_k(x; \Theta_k), \\
        0 ~~~ & \text{otherwise},
     \end{array}
\right.
\end{align*}
and its probability is:
\begin{align*}
\left\{
    \begin{array}{ll}
        \mathbb{P}(\Delta_{i,k} = 1) = w_k, \\
        \mathbb{P}(\Delta_{i,k} = 0) = 1-w_k.
     \end{array}
\right.
\end{align*}
Therefore, the log-likelihood can be written as:
\begin{align*}
& \ell(\Theta_1, \dots, \Theta_K) = \\
& \left\{
    \begin{array}{ll}
        \sum_{i=1}^n \log \big[w_1\, g_1(x_i; \Theta_1) \big]  \\
        ~~~~~~~~~~~~~~~~ \text{if } \Delta_{i,1} = 1 \text{ and } \Delta_{i,k} = 0 ~~\forall k \neq 1\\\\
        \sum_{i=1}^n \log \big[ w_2\, g_2(x_i; \Theta_2) \big]  \\
        ~~~~~~~~~~~~~~~~ \text{if } \Delta_{i,2} = 1 \text{ and } \Delta_{i,k} = 0 ~~\forall k \neq 2\\
        ~~~~~~~~ \vdots \\
        \sum_{i=1}^n \log \big[ w_K\, g_K(x_i; \Theta_K) \big]  \\
        ~~~~~~~~~~~~~~~~ \text{if } \Delta_{i,K} = 1 \text{ and } \Delta_{i,k} = 0 ~~\forall k \neq K
     \end{array}
\right.
\end{align*}
The above expression can be restated as:
\begin{align*}
\ell(\Theta_1, \dots, \Theta_K) = \sum_{i=1}^n \Bigg[ \sum_{k=1}^K \Delta_{i,k} \log \big( w_k g_k(x_i; \Theta_k) \big) \Bigg].
\end{align*}
The $\Delta_{i,k}$ here is the incomplete (missing) datum because we do not know whether it is $\Delta_{i,k}=0$ or $\Delta_{i,k}=1$ for $x_i$ and a specific $k$. Therefore, using the EM algorithm, we try to estimate it by its expectation.

The E-step in EM:
\begin{align*}
Q(\Theta_1, \dots, \Theta_K) = \sum_{i=1}^n \Bigg[ \sum_{k=1}^K &\, \mathbb{E}[\Delta_{i,k}|X,\Theta_1, \dots, \Theta_K] \\
&\times \log \big( w_k g_k(x_i; \Theta_k) \big) \Bigg].
\end{align*}
The $\Delta_{i,k}$ is either $0$ or $1$; therefore:
\begin{alignat*}{1}
\mathbb{E}[\Delta_{i,k}|X, & \Theta_1, \dots, \Theta_K] \\
&= 0 \times \mathbb{P}(\Delta_{i,k}=0|X, \Theta_1, \dots, \Theta_K) \\
& + 1 \times \mathbb{P}(\Delta_{i,k}=1|X, \Theta_1, \dots, \Theta_K) \\
& = \mathbb{P}(\Delta_{i,k}=1|X, \Theta_1, \dots, \Theta_K).
\end{alignat*}
According to Bayes rule (equation (\ref{equation_Bayes_3})), we have:
\begin{align*}
\mathbb{P}(&\Delta_{i,k}=1 |X, \Theta_1, \dots, \Theta_K) \\
& = \frac{\mathbb{P}(X, \Theta_1, \dots, \Theta_K, \Delta_{i,k}=1)}{\mathbb{P}(X; \Theta_1, \dots, \Theta_K)} \\
& = \frac{\mathbb{P}(X, \Theta_1, \dots, \Theta_K| \Delta_{i,k}=1)\,\mathbb{P}(\Delta_{i,k}=1)}{\sum_{k'=1}^K \mathbb{P}(X, \Theta_1, \dots, \Theta_K| \Delta_{i,k'}=1)\mathbb{P}(\Delta_{i,k'}=1)}.
\end{align*}
The marginal probability in the denominator is:
\begin{align*}
\mathbb{P}(X; \Theta_1, \dots, \Theta_K) = \sum_{k'=1}^K w_{k'}\, g_{k'}(x_i;\Theta_{k'}).
\end{align*}
Assuming that $\widehat{\gamma}_{i,k} := \mathbb{E}[\Delta_{i,k}|X, \Theta_1, \dots, \Theta_K]$ (called responsibility of $x_i$), we have:
\begin{align}\label{equation_gamma_multiMixture}
\widehat{\gamma}_{i,k} = \frac{\widehat{w}_k\, g_k(x_i;\Theta_k)}{\sum_{k'=1}^K \widehat{w}_{k'}\, g_{k'}(x_i;\Theta_{k'})},
\end{align}
and 
\begin{equation}
\begin{aligned}
Q(\Theta_1, \dots, \Theta_K) = \sum_{i=1}^n \sum_{k=1}^K \widehat{\gamma}_{i,k} \log \big( w_k g_k(x_i; \Theta_k) \big).
\end{aligned}
\end{equation}
Some simplification of $Q(\Theta_1, \dots, \Theta_K)$ will help in next step:
\begin{equation*}
\begin{aligned}
Q(\Theta_1, & \dots, \Theta_K) = \\
& \sum_{i=1}^n \sum_{k=1}^K \Big[ \widehat{\gamma}_{i,k} \log w_k + \widehat{\gamma}_{i,k} \log g_k(x_i; \Theta_k) \Big].
\end{aligned}
\end{equation*}

The M-step in EM:
\begin{align*}
& \widehat{\Theta}_k, \widehat{w}_k = \arg \max_{\Theta_k, w_k} Q(\Theta_1, \dots, \Theta_K, w_1, \dots, w_K), \\
& \text{subject to} ~~~ \sum_{k=1}^K w_k = 1.
\end{align*}
Note that the function $Q(\Theta_1, \dots, \Theta_K)$ is also a function of $w_1, \dots, w_K$ and that is why we wrote it as $Q(\Theta_1, \dots, \Theta_K, w_1, \dots, w_K)$.

The above problem is a constrained optimization problem:
\begin{equation*}
\begin{aligned}
& \underset{\Theta_k, w_k}{\text{maximize}}
& & Q(\Theta_1, \dots, \Theta_K, w_1, \dots, w_K), \\
& \text{subject to}
& & \sum_{k=1}^K w_k = 1,
\end{aligned}
\end{equation*}
which can be solved using Lagrange multiplier (see Section \ref{section_lagrangeMultiplier}):
\begin{equation*}
\begin{aligned}
\mathcal{L}(\Theta_1, &\dots, \Theta_K, w_1, \dots, w_K, \alpha) \\ 
& = Q(\Theta_1, \dots, \Theta_K, w_1, \dots, w_K) - \alpha \big(\sum_{k=1}^K w_k - 1\big) \\
& = \sum_{i=1}^n \sum_{k=1}^K \Big[ \widehat{\gamma}_{i,k} \log w_k + \widehat{\gamma}_{i,k} \log g_k(x_i; \Theta_k) \Big] \\
& ~~~~ - \alpha \big(\sum_{k=1}^K w_k - 1\big)
\end{aligned}
\end{equation*}
\begin{align}
& \frac{\partial \mathcal{L}}{\partial \Theta_k} = \sum_{i=1}^n \frac{\widehat{\gamma}_{i,k}}{g_k(x_i; \Theta_k)} \frac{\partial g_k(x_i; \Theta_k)}{\partial \Theta_k} \overset{\text{set}}{=} 0 \label{equation_derivative_theta_multiMixture}\\
& \frac{\partial \mathcal{L}}{\partial w_k} = \sum_{i=1}^n \frac{\widehat{\gamma}_{i,k}}{w_k} - \alpha  \overset{\text{set}}{=} 0 \implies w_k = \frac{1}{\alpha} \sum_{i=1}^n \gamma_{i,k} \nonumber \\
& \frac{\partial \mathcal{L}}{\partial \alpha} = \sum_{k=1}^K w_k - 1 \overset{\text{set}}{=} 0 \implies \sum_{k=1}^K w_k = 1 \nonumber \\
& \therefore ~~~~ \sum_{k=1}^K \frac{1}{\alpha} \sum_{i=1}^n \gamma_{i,k} = 1 \implies \alpha = \sum_{i=1}^n \sum_{k=1}^K \gamma_{i,k} \nonumber \\
& \therefore ~~~~ \widehat{w}_k = \frac{\sum_{i=1}^n \gamma_{i,k}}{\sum_{i=1}^n \sum_{k'=1}^K \gamma_{i,k'}} \label{equation_w_multiMixture}
\end{align}
Solving equations (\ref{equation_derivative_theta_multiMixture}) and (\ref{equation_w_multiMixture}) gives us the estimations $\widehat{\Theta}_k$ and $\widehat{w}_k$ (for $k \in \{1, \dots, K\}$) in every iteration.

The iterative algorithm for finding the parameters of the mixture of several distributions is shown in Algorithm \ref{algorithm_multiMixture}.

\SetAlCapSkip{0.5em}
\IncMargin{0.8em}
\begin{algorithm2e}[!t]
\DontPrintSemicolon
    \textbf{START:}
	\textbf{Initialize} $\widehat{\Theta}_1, \dots, \widehat{\Theta}_K$, $\widehat{w}_1, \dots, \widehat{w}_K$\;
	\While{not converged}{
	    \textit{// E-step in EM:}\;
	    \For{$i$ from $1$ to $n$}{
	        \For{$k$ from $1$ to $K$}{
                $\widehat{\gamma}_{i,k}  \gets $ equation (\ref{equation_gamma_multiMixture})\;
            }
        }
        \textit{// M-step in EM:}\;
        \For{$k$ from $1$ to $K$}{
    	    $\widehat{\Theta}_k  \gets $ equation (\ref{equation_derivative_theta_multiMixture})\;
    	    $\widehat{w}_k \gets $ equation (\ref{equation_w_multiMixture})\;
	    }
	    \textit{// Check convergence:}\;
	    Compare $\widehat{\Theta}_1, \dots, \widehat{\Theta}_K$, and $\widehat{w}_1, \dots, \widehat{w}_K$ with their values in previous iteration
	}
\caption{Fitting A Mixture of Several Distributions}\label{algorithm_multiMixture}
\end{algorithm2e}
\DecMargin{0.8em}

\subsubsection{Mixture of Several Gaussians}

Here, we consider a mixture of $K$ one-dimensional Gaussian distributions as an example for mixture of several continuous distributions. In this case, we have:
\begin{alignat*}{2}
& g_k(x; \mu_k, \sigma_k^2) &&= \frac{1}{\sqrt{2\pi\sigma_k^2}} \exp (-\frac{(x-\mu_k)^2}{2\sigma_k^2}) \\
& &&= \phi(\frac{x-\mu_k}{\sigma_k}), ~~~ \forall k \in \{1, \dots, K\}
\end{alignat*}
Therefore, equation (\ref{equation_mixture}) becomes:
\begin{equation}\label{equation_multiGaussians_general}
\begin{aligned}
f(x; \mu_1, \dots, \mu_K, \sigma_1^2, \dots, \sigma_K^2) = \sum_{k=1}^K w_k\, \phi(\frac{x-\mu_k}{\sigma_k}).
\end{aligned}
\end{equation}
The equation (\ref{equation_gamma_multiMixture}) becomes: 
\begin{align}\label{equation_gamma_multiGaussians}
\widehat{\gamma}_{i,k} = \frac{\widehat{w}_k\, \phi(\frac{x_i-\mu_k}{\sigma_k})}{\sum_{k'=1}^K \widehat{w}_{k'}\, \phi(\frac{x_i-\mu_{k'}}{\sigma_{k'}})}.
\end{align}
The $Q(\mu_1, \dots, \mu_K, \sigma_1^2, \dots, \sigma_K^2)$ is:
\begin{alignat*}{2}
Q(\mu_1,& \dots, \mu_K, \sigma_1^2, && \dots, \sigma_K^2) \\
& =\sum_{i=1}^n \sum_{k=1}^K \Big[ &&\widehat{\gamma}_{i,k} \log w_k + \widehat{\gamma}_{i,k} \big(\!-\frac{1}{2} \log (2\pi) \\
& && - \log \sigma_k -\frac{(x_i-\mu_k)^2}{2\sigma_k^2}\big) \Big].
\end{alignat*}
The Lagrangian is:
\begin{alignat*}{2}
\mathcal{L}(\mu_1, &\dots, \mu_K, \sigma_1^2, && \dots, \sigma_K^2, w_1, \dots, w_K, \alpha) \\ 
& =\sum_{i=1}^n \sum_{k=1}^K \Big[ && \widehat{\gamma}_{i,k} \log w_k + \widehat{\gamma}_{i,k} \big(\!-\frac{1}{2} \log (2\pi) \\
& && - \log \sigma_k -\frac{(x_i-\mu_k)^2}{2\sigma_k^2}\big) \Big] \\
& - \alpha \big(\sum_{k=1}^K w_k && - 1\big).
\end{alignat*}
Therefore:
\begin{align}
& \frac{\partial \mathcal{L}}{\partial \mu_k} = \sum_{i=1}^n \Big[\widehat{\gamma}_{i,k}\, (\frac{x_i-\mu_k}{\sigma_k^2}) \Big] \overset{\text{set}}{=} 0, \nonumber \\
& \implies \widehat{\mu}_k = \frac{\sum_{i=1}^n \widehat{\gamma}_{i,k}\, x_i}{\sum_{i=1}^n \widehat{\gamma}_{i,k}}, \label{equation_mu_multiGaussians}\\
& \frac{\partial \mathcal{L}}{\partial \sigma_k} = \sum_{i=1}^n \Big[\widehat{\gamma}_{i,k}\, (\frac{-1}{\sigma_k} + \frac{(x_i - \mu_k)^2}{\sigma_k^3}) \Big] \overset{\text{set}}{=} 0, \nonumber \\
& \implies \widehat{\sigma}_k^2 = \frac{\sum_{i=1}^n \widehat{\gamma}_{i,k}\, (x_i-\widehat{\mu}_k)^2}{\sum_{i=1}^n \widehat{\gamma}_{i,k}}, \label{equation_sigma_multiGaussians}
\end{align}
and $\widehat{w}_k$ is the same as equation (\ref{equation_w_multiMixture}).

Iteratively solving equations (\ref{equation_gamma_multiGaussians}), (\ref{equation_mu_multiGaussians}), (\ref{equation_sigma_multiGaussians}), and (\ref{equation_w_multiMixture}) using Algorithm (\ref{algorithm_multiMixture}) gives us the estimations for $\widehat{\mu}_1, \dots, \widehat{\mu}_K$, $\widehat{\sigma}_1, \dots, \widehat{\sigma}_K$, and $\widehat{w}_1, \dots, \widehat{w}_K$ in equation (\ref{equation_multiGaussians_general}).

\subsubsection{Multivariate Mixture of Gaussians}

The data might be multivariate ($\b{x} \in \mathbb{R}^d$) and the Gaussian distributions in the mixture model should be multi-dimensional in this case. We consider a mixture of $K$ multivariate Gaussian distributions. In this case, we have:
\begin{alignat*}{2}
& g_k(\b{x}; \,&&\b{\mu}_k, \b{\Sigma}_k) \\
& &&= \frac{1}{\sqrt{(2\pi)^d|\b{\Sigma}_k|}} \exp (-\frac{(\b{x}-\b{\mu}_k)^\top \b{\Sigma}_k^{-1} (\b{x}-\b{\mu}_k)}{2}) \\
& && \quad\quad\quad\quad\quad\quad\quad\quad\quad\quad\quad\quad~~~ \forall k \in \{1, \dots, K\},
\end{alignat*}
where $|\b{\Sigma}_k|$ is the determinant of $\b{\Sigma}_k$.

Therefore, equation (\ref{equation_mixture}) becomes:
\begin{equation}\label{equation_multiGaussians_general_multidimensional}
\begin{aligned}
f(\b{x}; \b{\mu}_1, \dots, \b{\mu}_K, \b{\Sigma}_1, \dots, \b{\Sigma}_K) = \sum_{k=1}^K w_k\, g_k(\b{x}; \b{\mu}_k, \b{\Sigma}_k).
\end{aligned}
\end{equation}
The equation (\ref{equation_gamma_multiMixture}) becomes: 
\begin{align}\label{equation_gamma_multiGaussians_multidimensional}
\widehat{\gamma}_{i,k} = \frac{\widehat{w}_k\, g_k(\b{x}_i; \b{\mu}_k, \b{\Sigma}_k)}{\sum_{k'=1}^K \widehat{w}_{k'}\, g_{k'}(\b{x}_i; \b{\mu}_{k'}, \b{\Sigma}_{k'})},
\end{align}
where $\b{x}_1, \dots, \b{x}_n \in \mathbb{R}^d$ and $\b{\mu}_1, \dots, \b{\mu}_K \in \mathbb{R}^d$ and $\b{\Sigma}_1, \dots, \b{\Sigma}_K \in \mathbb{R}^{d \times d}$ and $\widehat{w}_k \in \mathbb{R}$ and $\widehat{\gamma}_{i,k} \in \mathbb{R}$.

The $Q(\b{\mu}_1, \dots, \b{\mu}_K, \b{\Sigma}_1, \dots, \b{\Sigma}_K)$ is:
\begin{alignat*}{2}
Q(&\b{\mu}_1, \dots, \b{\mu}_K, &&\, \b{\Sigma}_1, \dots, \b{\Sigma}_K) \\
& =\sum_{i=1}^n \sum_{k=1}^K \Bigg[ &&\widehat{\gamma}_{i,k} \log w_k + \widehat{\gamma}_{i,k} \Big(\!\!-\frac{d}{2} \log (2\pi) \\
& && -\frac{1}{2} \log |\b{\Sigma}_k| \\
& && -\frac{1}{2} \textbf{tr}\big[ (\b{x}_i-\b{\mu}_k)^\top \b{\Sigma}_k^{-1} (\b{x}_i-\b{\mu}_k) \big]\Big) \Bigg],
\end{alignat*}
where $\textbf{tr}(.)$ denotes the trace of matrix. The trace is used here because $(\b{x}_i-\b{\mu}_k)^\top \b{\Sigma}_k^{-1} (\b{x}_i-\b{\mu}_k)$ is a scalar so it is equal to its trace.  

The Lagrangian is:
\begin{alignat*}{2}
\mathcal{L}(&\b{\mu}_1, \dots, \b{\mu}_K, &&\, \b{\Sigma}_1, \dots, \b{\Sigma}_K, w_1, \dots, w_K, \alpha) \\ 
& =\sum_{i=1}^n \sum_{k=1}^K \Bigg[ &&\widehat{\gamma}_{i,k} \log w_k + \widehat{\gamma}_{i,k} \Big(\!\!-\frac{d}{2} \log (2\pi) \\
& && -\frac{1}{2} \log |\b{\Sigma}_k| \\
& && -\frac{1}{2} \textbf{tr}\big[ (\b{x}_i-\b{\mu}_k)^\top \b{\Sigma}_k^{-1} (\b{x}_i-\b{\mu}_k) \big]\Big) \Bigg] \\
& - \alpha \big(\sum_{k=1}^K w_k && - 1\big).
\end{alignat*}
Therefore:
\begin{align}
& \frac{\partial \mathcal{L}}{\partial \b{\mu}_k} = \sum_{i=1}^n \Big[\widehat{\gamma}_{i,k}\, \b{\Sigma}_k^{-1} (\b{x}_i-\b{\mu}_k) \Big] \overset{\text{set}}{=} \b{0} \in \mathbb{R}^d, \nonumber \\
& \overset{(a)}{\implies} \sum_{i=1}^n \Big[\widehat{\gamma}_{i,k}\, (\b{x}_i-\b{\mu}_k) \Big] = \b{0}, \nonumber \\
& \implies \widehat{\b{\mu}}_k = \frac{\sum_{i=1}^n \widehat{\gamma}_{i,k}\, \b{x}_i}{\sum_{i=1}^n \widehat{\gamma}_{i,k}} \in \mathbb{R}^{d}, \label{equation_mu_multiGaussians_multidimensional}\\
& \frac{\partial \mathcal{L}}{\partial \b{\Sigma}_k} \overset{(b)}{=} \sum_{i=1}^n \Big[\widehat{\gamma}_{i,k}\, (\frac{-1}{2} \b{\Sigma}_k \nonumber \\
& ~~~~~~~~~~~~ + \frac{1}{2}(\b{x}_i - \b{\mu}_k)(\b{x}_i - \b{\mu}_k)^\top \Big] \overset{\text{set}}{=} \b{0} \in \mathbb{R}^{d \times d}, \nonumber \\
& \implies \b{\Sigma}_k \sum_{i=1}^n \widehat{\gamma}_{i,k} = \sum_{i=1}^n \widehat{\gamma}_{i,k} (\b{x}_i - \b{\mu}_k)(\b{x}_i - \b{\mu}_k)^\top, \nonumber \\
& \implies \widehat{\b{\Sigma}}_k = \frac{\sum_{i=1}^n \widehat{\gamma}_{i,k} (\b{x}_i - \b{\mu}_k)(\b{x}_i - \b{\mu}_k)^\top}{\sum_{i=1}^n \widehat{\gamma}_{i,k}} \in \mathbb{R}^{d \times d}, \label{equation_sigma_multiGaussians_multidimensional}
\end{align}
and $\widehat{w}_k \in \mathbb{R}$ is the same as equation (\ref{equation_w_multiMixture}).
In above expressions, $(a)$ is because $\b{\Sigma}_k^{-1} \neq \b{0} \in \mathbb{R}^{d \times d}$ is not dependent on $i$ and can be left factored out of the summation (note that $\widehat{\gamma}_{i,k}$ is a scalar), and $(b)$ is because $\frac{\partial}{\partial \b{\Sigma}_k} \log |\b{\Sigma}_k| = \b{\Sigma}_k$ and $\textbf{tr}\big[ (\b{x}_i-\b{\mu}_k)^\top \b{\Sigma}_k^{-1} (\b{x}_i-\b{\mu}_k) \big] = \textbf{tr}\big[ \b{\Sigma}_k^{-1} (\b{x}_i-\b{\mu}_k) (\b{x}_i-\b{\mu}_k)^\top \big]$ and $\frac{\partial}{\partial \b{\Sigma}_k} \textbf{tr}\big[ \b{\Sigma}_k^{-1} \b{A} \big] = - \b{A}$.

Iteratively solving equations (\ref{equation_gamma_multiGaussians_multidimensional}), (\ref{equation_mu_multiGaussians_multidimensional}), (\ref{equation_sigma_multiGaussians_multidimensional}), and (\ref{equation_w_multiMixture}) using Algorithm (\ref{algorithm_multiMixture}) gives us the estimations for $\widehat{\b{\mu}}_1, \dots, \widehat{\b{\mu}}_K$, $\widehat{\b{\Sigma}}_1, \dots, \widehat{\b{\Sigma}}_K$, and $\widehat{w}_1, \dots, \widehat{w}_K$ in equation (\ref{equation_multiGaussians_general_multidimensional}). 
The multivariate mixture of Gaussians is also mentioned in \cite{lee2012algorithms}.
Moreover, note that the mixture of Gaussians is also referred to as Gaussian Mixture Models (GMM) in the literature.

\subsubsection{Mixture of Several Poissons}

Here, we consider a mixture of $K$ Poisson distributions as an example for mixture of several discrete distributions. In this case, we have:
\begin{align*}
g_k(x; \lambda_k) = \frac{e^{-\lambda_k} \lambda_k^{x}}{x!},
\end{align*}
therefore, equation (\ref{equation_mixture}) becomes:
\begin{equation}\label{equation_multiPoissons_general}
\begin{aligned}
f(x; \lambda_1, \dots, \lambda_K) = \sum_{k=1}^K w_k\, \frac{e^{-\lambda_k} \lambda_k^{x}}{x!}.
\end{aligned}
\end{equation}
The equation (\ref{equation_gamma_multiMixture}) becomes:
\begin{align}\label{equation_gamma_multiPoissons}
\widehat{\gamma}_{i,k} = \frac{\widehat{w}_k\, (\frac{e^{-\widehat{\lambda}_k} \widehat{\lambda}_k^{x_i}}{x_i!})}{\sum_{k'=1}^K \widehat{w}_{k'}\, (\frac{e^{-\widehat{\lambda}_{k'}} \widehat{\lambda}_{k'}^{x_i}}{x_i!})}.
\end{align}
The $Q(\lambda_1, \dots, \lambda_K)$ is:
\begin{alignat*}{2}
Q(\lambda_1, \dots, \lambda_K) = &\sum_{i=1}^n \sum_{k=1}^K \Big[ \widehat{\gamma}_{i,k} \log w_k 
\\& + \widehat{\gamma}_{i,k} (-\lambda_k + x_i \log \lambda_k - \log x_i!) \Big].
\end{alignat*}
The Lagrangian is:
\begin{alignat*}{2}
\mathcal{L}(\lambda_1, &\dots, \lambda_K, w_1, &&\dots, w_K, \alpha) \\ 
& =\sum_{i=1}^n \sum_{k=1}^K \Big[ && \widehat{\gamma}_{i,k} \log w_k \\
& && + \widehat{\gamma}_{i,k} (-\lambda_k + x_i \log \lambda_k - \log x_i!) \Big] \\
& - \alpha \big(\sum_{k=1}^K w_k && - 1\big).
\end{alignat*}

Therefore:
\begin{align}
& \frac{\partial \mathcal{L}}{\partial \lambda_k} = \sum_{i=1}^n \Big[\widehat{\gamma}_{i,k} (-1+\frac{x_i}{\lambda_k}) \Big] \overset{\text{set}}{=} 0, \nonumber \\
& \implies \widehat{\lambda}_k = \frac{\sum_{i=1}^n \widehat{\gamma}_{i,k}\, x_i}{\sum_{i=1}^n \widehat{\gamma}_{i,k}}, \label{equation_lambda_multiPoissons}
\end{align}
and $\widehat{w}$ is the same as equation (\ref{equation_w_multiMixture}).

Iteratively solving equations (\ref{equation_gamma_multiPoissons}), (\ref{equation_lambda_multiPoissons}), and (\ref{equation_w_multiMixture}) using Algorithm (\ref{algorithm_multiMixture}) gives us the estimations for $\widehat{\lambda}_1, \dots, \widehat{\lambda}_K$, and $\widehat{w}_1, \dots, \widehat{w}_K$ in equation (\ref{equation_multiPoissons_general}).

\section{Using Mixture Distribution for Clustering}\label{section_clustering}

Mixture distributions have a variety of applications including clustering. Assuming that the number of clusters, denoted by $K$, is known, the cluster label of a point $x_i$ ($i \in \{1, \dots, n\}$) is determined as:
\begin{align}
\text{label of } x_i \gets \arg \max_k g_k(x_i; \Theta_k),
\end{align}
where $g_k(x_i; \Theta_k)$ is the $k$-th distribution fitted to data $x_1, \dots, x_n$. In other words, where $f(x; \Theta_1, \dots, \Theta_K) = \sum_{k=1}^K w_k\, g_k(x; \Theta_k)$ is the fitted mixture distribution to data. The reason of why this clustering works is that the density/mass function which has produced that point with higher probability can be the best candidate for the cluster of that point. 
This method of clustering is referred to as ``model-based clustering'' in literature \cite{fraley1998many,fraley2002model}. 

\section{Simulations}\label{section_simulations}

In this section, we do some simulations on fitting a mixture of densities in both continuous and discrete cases. For continuous cases, a mixture of three Gaussians and for discrete cases, a mixture of three Poissons are simulated.

\subsection{Mixture of Three Gaussians}

\begin{figure}[!t]
\centering
\includegraphics[width=3.25in]{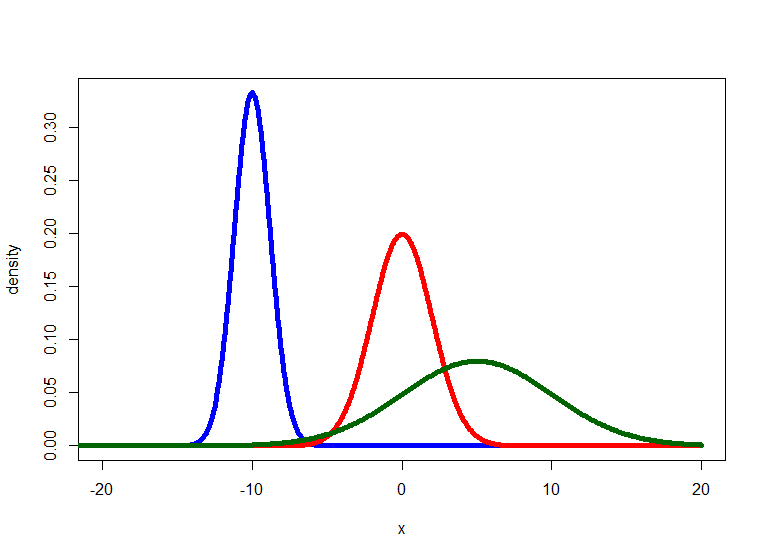}
\caption{The original probability density functions from which the sample is drawn. The mixture includes three different Gaussians showed in blue, red, and green colors.}
\label{figure_gaussian_original_densities}
\end{figure}

A sample with size $n=2200$ from three distributions is randomly generated for this experiment:
\begin{align*}
& \phi(\frac{x - \mu_1}{\sigma_1}) = \phi(\frac{x + 10}{1.2}), \\
& \phi(\frac{x - \mu_2}{\sigma_2}) = \phi(\frac{x - 0}{2}), \\
& \phi(\frac{x - \mu_3}{\sigma_3}) = \phi(\frac{x - 5}{5}).
\end{align*}
For having generality, the size of subset of sample generated from the three densities are different, i.e., $700$, $1000$, and $500$. The three densities are shown in Fig. \ref{figure_gaussian_original_densities}. 

\begin{figure}[!t]
\centering
\includegraphics[width=3.25in]{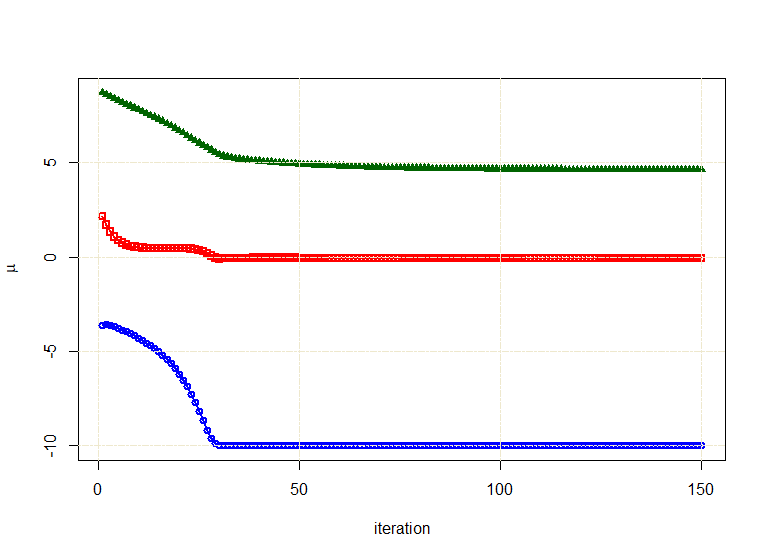}
\caption{The change and convergence of $\mu_1$ (shown in blue), $\mu_2$ (shown in red), and $\mu_3$ (shown in green) over the iterations.}
\label{figure_gaussian_mu}
\end{figure}

Applying Algorithm \ref{algorithm_multiMixture} and using equations (\ref{equation_gamma_multiGaussians}), (\ref{equation_mu_multiGaussians}), (\ref{equation_sigma_multiGaussians}), and (\ref{equation_w_multiMixture}) for mixture of $K=3$ Gaussians gives us the estimated values for the parameters:
\begin{alignat*}{4}
& \mu_1 = -9.99, ~ &&\sigma_1 = 1.17, ~ &&&w_1 = 0.317 \\
& \mu_2 = -0.05, ~ &&\sigma_2 = 1.93, ~ &&&w_2 = 0.445 \\
& \mu_3 = 4.64, ~ &&\sigma_3 = 4.86, ~ &&&w_3 = 0.237 
\end{alignat*}
Comparing the estimations for $\mu_1, \mu_2, \mu_3$ and $\sigma_1, \sigma_2, \sigma_3$ with those in original densities from which data were generated verifies the correctness of the estimations.

\begin{figure}[!t]
\centering
\includegraphics[width=3.25in]{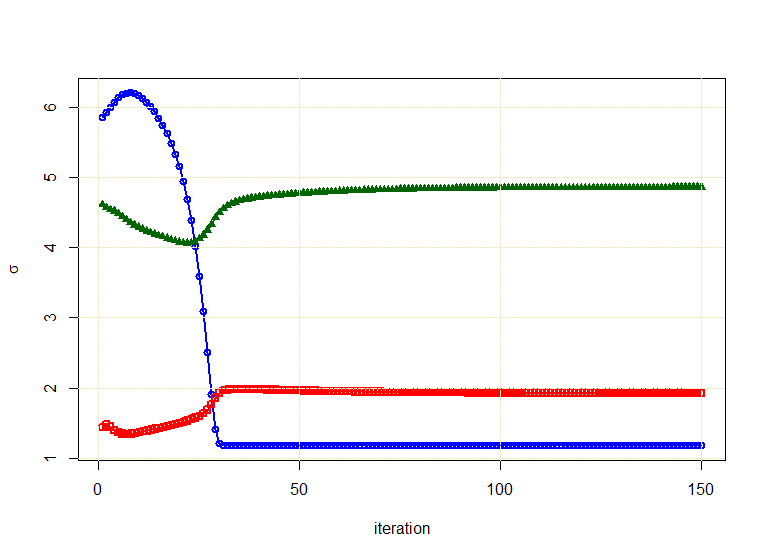}
\caption{The change and convergence of $\sigma_1$ (shown in blue), $\sigma_2$ (shown in red), and $\sigma_3$ (shown in green) over the iterations.}
\label{figure_gaussian_sigma}
\end{figure}

\begin{figure}[!t]
\centering
\includegraphics[width=3.25in]{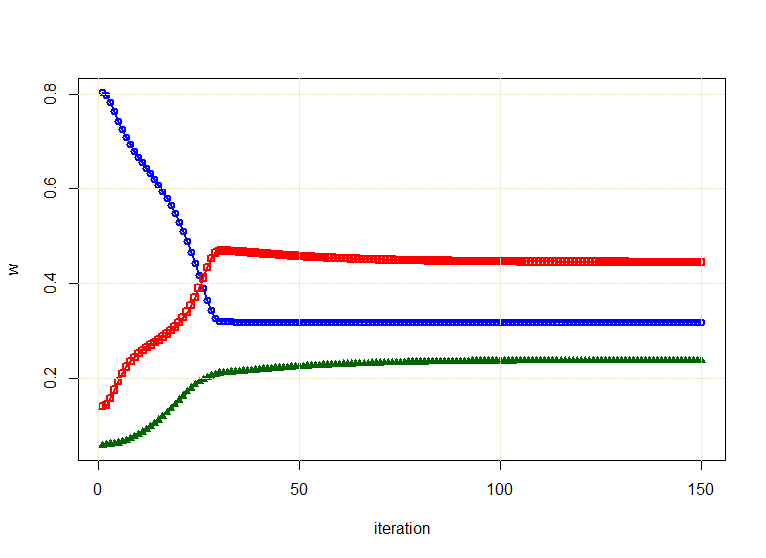}
\caption{The change and convergence of $w_1$ (shown in blue), $w_2$ (shown in red), and $w_3$ (shown in green) over the iterations.}
\label{figure_gaussian_weights}
\end{figure}

The progress of the parameters $\mu_k$, $\sigma_k$, and $w_k$ through the iterations until convergence are shown in figures \ref{figure_gaussian_mu}, \ref{figure_gaussian_sigma}, and \ref{figure_gaussian_weights}, respectively.

\begin{figure}[!t]
\centering
\includegraphics[width=3.25in]{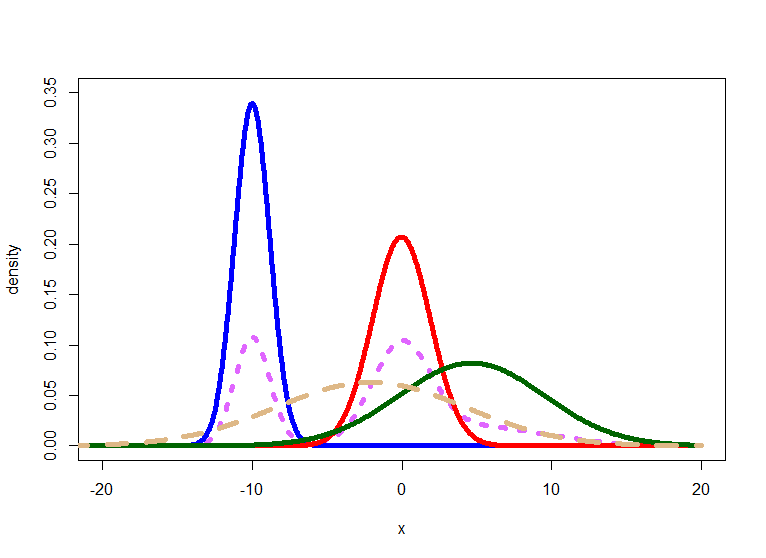}
\caption{The estimated probability density functions. The estimated mixture includes three different Gaussians showed in blue, red, and green colors. The dashed purple density is the weighted summation of these three densities, i.e., $\sum_{k=1}^3 w_k \phi(\frac{x-\mu_k}{\sigma_k})$. The dashed brown density is the fitted density whose parameters are estimated by MLE.}
\label{figure_gaussian_fitted_distributions_2}
\end{figure}

Note that for setting initial values of parameters in mixture of Gaussians, one reasonable option is:
\begin{align}
& \text{range} \gets \max_i(x_i) - \min_i(x_i), \nonumber \\
& \mu_k^{(0)} \sim U(\min_i(x_i), \max_i(x_i)), \\
& \sigma_k^{(0)} \sim U(0, \text{range} / 6), \\
& w_k^{(0)} \sim U(0, 1),
\end{align}
where $U(\alpha, \beta)$ is continuous uniform distribution in range $(\alpha, \beta)$. This initialization makes sense because in normal distribution, the mean belongs to the range of data and $99\%$ of data falls in range $(\mu-3\sigma, \mu+3\sigma)$; therefore, the spread of data is roughly $6\sigma$. 
In the experiment of this section, the mentioned initialization is utilized. 

The fitted densities and the mixture distribution are depicted in Fig. \ref{figure_gaussian_fitted_distributions_2}. Comparing this figure with Fig. \ref{figure_gaussian_original_densities} verifies the correct estimation of the three densities. Figure \ref{figure_gaussian_fitted_distributions_2} also shows the mixture distribution, i.e., the weighted summation of the estimated densities. 

Moreover, for the sake of better comparison, only one distribution is also fitted to data using MLE. The MLE estimation of parameters are $\widehat{\mu}^{(mle)} = \bar{x} = (1/n) \sum_{i=1}^n x_i$ and $\widehat{\sigma}^{(mle)} = (1/n) \sum_{i=1}^n (x_i - \bar{x})^2$. This fitted distribution is also depicted in Fig. \ref{figure_gaussian_fitted_distributions_2}. We can see that this poor estimation has not captured the multi-modality of data in contrast to the estimated mixture distribution.

\subsection{Mixture of Three Poissons}

\begin{figure}[!t]
\centering
\includegraphics[width=3.25in]{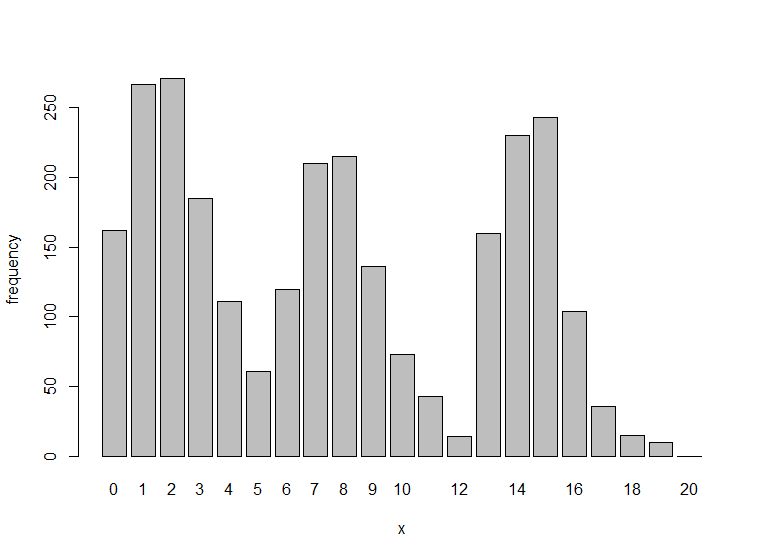}
\caption{The frequency of the discrete data sample.}
\label{figure_poisson_original_data}
\end{figure}

A sample with size $n=2666$ is made (see Table \ref{table_poisson_original_data}) for the experiment where the frequency of data, displayed in Fig \ref{figure_poisson_original_data}, shows that data are almost sampled from a mixture of three Poissons. 
 
\begin{table}[!t]
\setlength\extrarowheight{5pt}
\centering
\scalebox{0.65}{    
\begin{tabular}{c || c | c | c | c | c | c | c | c | c | c | c}
$x$ & 0 & 1 & 2 & 3 & 4 & 5 & 6 & 7 & 8 & 9 & 10 \\
\hline
frequency & 162 & 267 & 271 & 185 & 111 & 61 & 120 & 210 & 215 & 136 & 73 \\
\hline
\hline
$x$ & 11 & 12 & 13 & 14 & 15 & 16 & 17 & 18 & 19 & 20 & \\
\hline
frequency & 43 & 14 & 160 & 230 & 243 & 104 & 36 & 15 & 10 & 0 \\
\end{tabular}%
}
\caption{The discrete data for simulation of fitting mixture of Poissons.}
\label{table_poisson_original_data}
\end{table}

\begin{figure}[!t]
\centering
\includegraphics[width=3.25in]{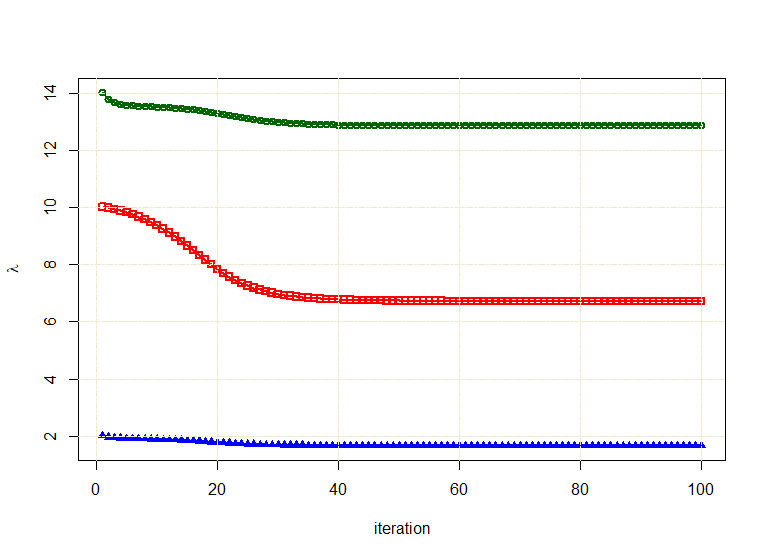}
\caption{The change and convergence of $\lambda_1$ (shown in blue), $\lambda_2$ (shown in red), and $\lambda_3$ (shown in green) over the iterations.}
\label{figure_poisson_mu}
\end{figure}

\begin{figure}[!t]
\centering
\includegraphics[width=3.25in]{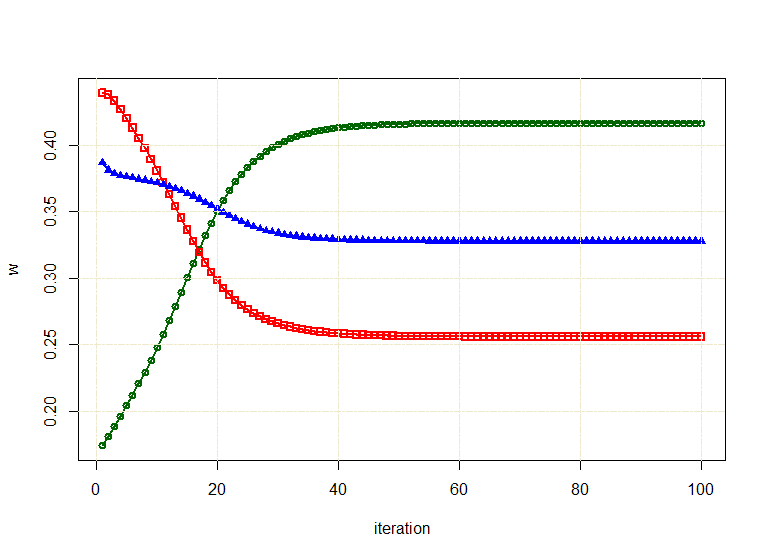}
\caption{The change and convergence of $w_1$ (shown in blue), $w_2$ (shown in red), and $w_3$ (shown in green) over the iterations.}
\label{figure_poisson_weights}
\end{figure}

Applying Algorithm \ref{algorithm_multiMixture} and using equations (\ref{equation_gamma_multiPoissons}), (\ref{equation_lambda_multiPoissons}), and (\ref{equation_w_multiMixture}) for mixture of $K=3$ Poissons gives us the estimated values for the parameters:
\begin{alignat*}{3}
& \lambda_1 = 1.66, ~ &&w_1 = 0.328 \\
& \lambda_2 = 6.72, ~ &&w_2 = 0.256 \\
& \lambda_3 = 12.85, ~ &&w_3 = 0.416 
\end{alignat*}
Comparing the estimations for $\lambda_1, \lambda_2, \lambda_3$ with Fig. \ref{figure_poisson_original_data} verifies the correctness of the estimations.
The progress of the parameters $\lambda_k$ and $w_k$ through the iterations until convergence are shown in figures \ref{figure_poisson_mu} and \ref{figure_poisson_weights}, respectively. 

For setting initial values of parameters in mixture of Poissons, one reasonable option is:
\begin{align}
& \lambda_k^{(0)} \sim U(\min_i(x_i), \max_i(x_i)), \\
& w_k^{(0)} \sim U(0, 1).
\end{align}
The reason of this initialization is that the MLE estimation of $\lambda$ is $\widehat{\lambda}^{(mle)} = \bar{x} = (1/n) \sum_{i=1}^n x_i$ which belongs to the range of data.
This initialization is used in this experiment.

\begin{figure}[!t]
\centering
\includegraphics[width=3.25in]{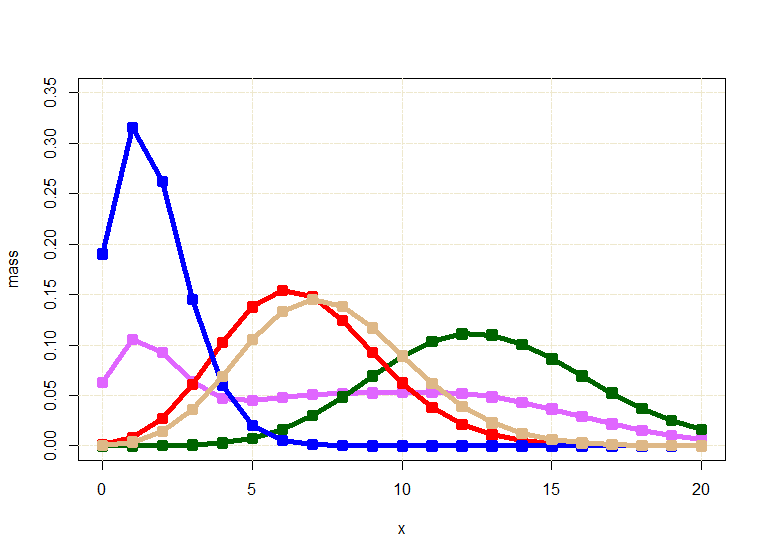}
\caption{The estimated probability mass functions. The estimated mixture includes three different Poissons showed in blue, red, and green colors. The purple density is the weighted summation of these three densities, i.e., $\sum_{k=1}^3 w_k \frac{e^{-\lambda_k} \lambda^k}{x!}$. The brown density is the fitted density whose parameter is estimated by MLE.}
\label{figure_poisson_fitted_distributions_2}
\end{figure}

The fitted mass functions and the mixture distribution are depicted in Fig. \ref{figure_poisson_fitted_distributions_2}. Comparing this figure with Fig. \ref{figure_poisson_original_data} verifies the correct estimation of the three mass functions. The mixture distribution, i.e., the weighted summation of the estimated densities, is also shown in Fig. \ref{figure_poisson_fitted_distributions_2}.

For having better comparison, only one mass function is also fitted to data using MLE. For that, the parameter $\lambda$ is estimated using $\widehat{\lambda}^{(mle)} = \bar{x} = (1/n) \sum_{i=1}^n x_i$. This fitted distribution is also depicted in Fig. \ref{figure_poisson_fitted_distributions_2}. Again, the poor performance of this single mass function in capturing the multi-modality is obvious. 

\section{Conclusion}\label{section_conclusions}

In this paper, a simple-to-understand and step-by-step tutorial on fitting a mixture distribution to data was proposed. The assumption was the prior knowledge on calculus and basic linear algebra. For more clarification, fitting two distributions was primarily introduced and then it was generalized to $K$ distributions. Fitting mixture of Gaussians and Poissons were also mentioned as examples for continuous and discrete cases, respectively. Simulations were also shown for more clarification.

\section*{Acknowledgment}
The authors hugely thank Prof. Mu Zhu for his great course ``Statistical Concepts for Data Science''. This great course partly covered the materials mentioned in this tutorial paper. 


\bibliography{References}
\bibliographystyle{icml2016}

\end{document}